\documentclass[aps,prx,twocolumn,superscriptaddress,showpacs,amsmath,amssymb,longbibliography]{revtex4-1}
\usepackage{mathtools}
\usepackage{amsfonts}
\usepackage{amssymb}
\usepackage{graphicx}
\usepackage{color}

\usepackage{bm}
\usepackage{braket}
\usepackage{gensymb}

\usepackage{hyperref}

\newcommand{\calT}{\mathcal{T}}

\DeclareMathOperator{\Tr}{Tr}

\begin{document}

\title{Generalized $f$-Sum Rules and Kohn formulas on Non-linear Conductivities}

\author{Haruki Watanabe}
\affiliation{Department of Applied Physics, University of Tokyo, Tokyo 113-8656, Japan.}

\author{Masaki Oshikawa}
\affiliation{Institute for Solid State Physics, University of Tokyo, Kashiwa 277-8581, Japan.}
\affiliation{Kavli Institute for the Physics and Mathematics of the Universe (WPI), University of Tokyo, Kashiwa 277-8583, Japan}

\begin{abstract}
The $f$-sum rule and the Kohn formula are well-established general constraints on the electric conductivity in quantum many-body systems.
We present their generalization to non-linear conductivities at all orders of the response in a unified manner,
by considering two limiting quantum time-evolution processes: a quench process and an adiabatic process.
Our generalized formulas are valid in any stationary state, including the ground state and finite temperature Gibbs states, regardless of the details of the system such as the specific form of the kinetic term, the strength of the many-body interactions, or the presence of disorders.
\end{abstract}

\maketitle

\clearpage

\section{Introduction}
Understanding of dynamical responses of a quantum many-body system is
not only theoretically interesting but is also essential for bridging
theory and experiment, as many experiments measure dynamical responses.
Linear responses have been best understood, thanks to the general
framework of linear response theory~\cite{Kubo,Nakano,Kubo-StatPhysII}.
Many experiments can be actually well
described in terms of linear responses.
On the other hand, there is a renewed strong interest in nonlinear
responses recently,
thanks to new theoretical ideas, powerful numerical methods, and
developments in experimental techniques such as powerful laser sources
which enable us to probe highly nonlinear responses.  For example,
``shift current,'' which is a DC current induced by AC electric field as
a higher order effect, has been studied
vigorously~\cite{PhysRevB.23.5590,SipeShkrebtii2000,YoungRappe2012,MorimotoNagaosa2016,Fregoso2017,Yang-shift2017,MorimotoNagaosa2018}.

Yet, theoretical computations of dynamical responses are generally challenging,
often even for linear responses and more so for nonlinear ones.
Therefore it is useful to obtain general constraints on dynamical
responses, including their relations to \emph{static} quantities which are
easier to calculate.

The ``$f$-sum rule'' and the the ``Kohn formula" of the linear electric conductivity are typical
and well-known examples of such constraints.
They have played an indispensable role in many applications,
and their importance is well established~\cite{Pines-EE_Solids,Resta_2018}.
To introduce them, let us consider the simplest case of the uniform component ($\vec{q}=0$ Fourier component)
of the linear AC conductivity defined as
\begin{equation}
 j_i(\omega) = \sum_j\sigma_{i}^j(\omega) E_j(\omega),
\end{equation}
where $i,j$ are indices for spatial directions, $j_i(\omega)=j_i(-\omega)^*$ is the uniform electric current, and $E_j(\omega)=E_j(-\omega)^*$ is the uniform electric field.

The $f$-sum rule is a constraint on the frequency integral $\int_{-\infty}^{\infty}d\omega\,\sigma_{i}^j(\omega)$.   In condensed matter physics, the typical Hamiltonian has the form $\hat{H} = \hat{K} + \hat{I}$, where $\hat{K}$ is the kinetic energy (including the chemical potential term)
which is bilinear in particle creation/annihilation operators,
and $\hat{I}$ is the density-density interaction energy.  For the standard kinetic term in non-relativistic quantum mechanics in the continuum
$
\hat{K} = \int d^dr\,\hat{c}^\dagger_{\vec{r}}
\big[ - (\vec{\nabla}^2/2m) - \mu \big]
\hat{c}_{\vec{r}},
$
the original form of the $f$-sum rule is known as
\begin{equation}
  \int_{-\infty}^{\infty} \frac{d\omega}{2\pi} \sigma_{i}^j(\omega)
   = \delta_{ij}\frac{\rho}{2m}.
\label{eq.f-sum_NR}
\end{equation}
The right-hand side is determined by the electron mass $m$ and the
electron density $\rho$, and is a completely static quantity.  (Throughout the text we set $e=\hbar=1$.)
For more general models of the form $\hat{H} = \hat{K} + \hat{I}$, the
$f$-sum rule still holds although with a modified right-hand
side~\cite{Bari-fsum,Sadataka-fsum,Izuyama1973,Maldague1977,BaeriswylCarmeloLuther1986,LimtragoolPhillips2017,Hazra2018}.

The Kohn formula~\cite{Kohn1964} is an analytic expression of the Drude weight, also called the charge stiffness, that characterizes the ballistic transport of the system.
The Drude weight is formally defined by $\mathcal{D}_i^j=\lim_{\omega\to0}\omega\text{Im}\,\sigma_{i}^j(\omega)$. In other words, it appears in $\sigma_{i}^j(\omega)$ as
\begin{equation}
\sigma_{i}^j(\omega)=\frac{i}{\omega+i\eta}\mathcal{D}_i^j + \dots,
\label{eq.defDrude}
\end{equation}
where  $\eta>0$ is an infinitesimal convergent parameter and the dots denote terms regular around $\omega=0$.  (Our definition of $\mathcal{D}_i^j$ contains an additional factor of $2$ as compared to the standard convention in the literature.)
The Kohn formula gives the Drude weight
$\mathcal{D}_i^j$ at zero temperature in terms of the curvature of the ground state energy as a function of the twist of the boundary condition. The formula was extended to a finite temperature in Ref.~\onlinecite{PhysRevLett.74.972}. Its validity and subtlety in application to many-body systems have been investigated in Hubbard chains~\cite{PhysRevB.43.13660,ShastrySutherland-twisted,RigolShastry-Drude,PhysRevLett.74.972,PhysRevB.83.035115} and in Heisenberg spin chains~\cite{PhysRevLett.65.1833,ShastrySutherland-twisted,PhysRevLett.82.1764,PhysRevB.83.035115,ZotosJPSJ,BentzJPSJ}.

The main result of this work is the generalization of the $f$-sum rule and the Kohn formula on the linear conductivity, summarized above, to an infinite series of formulas on nonlinear conductivities at arbitrary orders.
Although nonlinear $f$-sum rules of general response functions have been formulated in Ref.~\cite{Shimizu}, the formulation there is not directly applicable to the $f$-sum rule of the optical conductivity. The results in Ref.~\cite{Shimizu} and its subsequent works~\cite{ShimizuYuge1,ShimizuYuge2} partially overlap ours, but our results are more general in several aspects. (See Ref.~\cite{derivation} for a more detailed comparison.)
Conventionally, the $f$-sum rule and the Drude weight are
formulated in the frequency space as in
Eqs.~\eqref{eq.f-sum_NR} and~\eqref{eq.defDrude}.
However, it is illuminating, and indeed useful as we demonstrate below,
to formulate them in terms of the real time response
of the current to the applied electric field.
The integral over the frequency for the $f$-sum rule
corresponds to the instantaneous response, and the singularity at zero
frequency which gives the Drude weight corresponds to the response
after an infinitely long time.
In fact, considering a very similar process of application of an
electric field pulse both in the quantum quench  (zero time) limit
and in the adiabatic (infinite time) limit,
we obtain the nonlinear generalizations of the $f$-sum rule
and the Kohn formula, respectively.  A similar idea has been utilized in the discussion of the Drude weight at the linear order earlier~\cite{Oshikawa-Drude,*Oshikawa-Drude-Erratum}.
The present approach allows us to treat the linear and nonlinear conductivities, and the $f$-sum rule and Drude weight, on the same footing in a unified framework.
Our results are quite general and not limited to the Hamiltonians of the form $\hat{H}=\hat{K}+\hat{I}$.
These results hold in any steady state including the ground state and in equilibrium at a finite temperature.

The remainder of this paper is organized as follows.
The setup and the main results of our study are summarized in Sec.~\ref{sec:setup}.
A simple proof of our claims is presented in Sec.~\ref{sec:proof1}.
Several examples are discussed in Sec.~\ref{example}.
The concluding remarks are in Sec.~\ref{sec:discussion}.

\section{Summary of results}
\label{sec:setup}
\subsection{Setup}
We consider a general system of many quantum particles.
To demonstrate our main claim in a simple setting, let us assume the $d$-dimensional cubic lattice and focus on the uniform component of the electric current induced by a uniform electric field.
The system size $V$ and the boundary condition can be chosen arbitrarily. We do not require any spatial symmetry such as the translation invariance
or the rotation symmetry.

The Hamiltonian of the system is written in terms of
creation and annihilation operators
$\hat{c}_{\vec{r}\alpha}^\dagger$,
$\hat{c}_{\vec{r}\alpha}$ ($\alpha$ labels the internal degrees of
freedom) defined on each point $\vec{r}$.
We allow any number of creation and annihilation operators to appear in
a single term in the Hamiltonian, representing correlated hopping, pair
hopping, ring exchange, and so on. Thus our Hamiltonian \emph{does not} necessarily take the form $\hat{H} = \hat{K} + \hat{I}$.
We still assume that all the hoppings and
interactions are short-ranged and U(1) symmetric.

We describe the electric field via the time-dependence of the U(1) vector potential $\vec{A}(t)=(A_x(t),A_y(t),\dots)$ while setting the scaler potential to be $0$.
In order to discuss the uniform electric field, we assume that every link in the $i$-th direction has the same value $A_i(t)$ ($i=x,y,\dots$).
The Hamiltonian $\hat{H}(\vec{A}(t))$ then depends on $t$ through $\vec{A}(t)$.  We set $\vec{A}(t)=0$ for $t\leq0$ and continuously turn it on for $t>0$.
The resulting electric field is
\begin{equation}
\vec{E}(t) \equiv\frac{d \vec{A}(t)}{dt}
\label{defE}
\end{equation}
(To avoid negative signs, we use the sign convention opposite to the standard definition.)
The U(1) symmetry of the Hamiltonian enables us to identify the
current density $\hat{\vec{j}}\equiv(\hat{j}_x,\hat{j}_y,\dots)$
averaged over the entire system:
\begin{equation}
\hat{j}_i(\vec{A})\equiv\frac{1}{V}\frac{\partial \hat{H}(\vec{A})}{\partial A_i}.\label{defj}
\end{equation}

Suppose that the system is described by a stationary state at $t=0$:
\begin{equation}
\hat{\rho}(0)=\sum_n\rho_n|n(\vec{0})\rangle\langle n(\vec{0})|,\quad\sum_n\rho_n=1.\label{initial}
\end{equation}
Here $|n(\vec{0})\rangle$ is the $n$-th eigenstate of the unperturbed Hamiltonian $\hat{H}(\vec{0})$ with the energy eigenvalue $\mathcal{E}_n(\vec{0})$. For example, the Gibbs state with an inverse temperature $\beta$ is given by $\rho_n=e^{-\beta \mathcal{E}_n(\vec{0})}/Z$ ($Z\equiv\sum_ne^{-\beta \mathcal{E}_n(\vec{0})}$).

The evolution of the system for $t\geq0$ is described by the time-evolution operator $\hat{S}(t)$ defined by
\begin{align}
\frac{d\hat{S}(t)}{dt}=-i \hat{H}(\vec{A}(t))\hat{S}(t),\quad \hat{S}(0)=1.\label{dS}
\end{align}
The expectation value of an operator $\hat{O}$ at time $t\geq0$ is then given by
\begin{align}
\langle \hat{O} \rangle_t \equiv\Tr[\hat{O}\hat{\rho}(t)],\quad \hat{\rho}(t)=\hat{S}(t)\hat{\rho}(0)\hat{S}(t)^\dagger.\label{rhot}
\end{align}

The linear and nonlinear conductivities in real time are defined as the response of the current density
\begin{equation}
j_i(t)\equiv\langle\hat{j}_{i}(\vec{A}(t))\rangle_t=\frac{1}{V}\Big\langle\frac{\partial \hat{H}(\vec{A})}{\partial A_i}\Big|_{\vec{A}=\vec{A}(t)}\Big\rangle_t\label{defj2}
\end{equation}
towards the applied electric field:
\begin{align}
&j_i(t)-j_i(0)=\sum_{N=1}^\infty\frac{1}{N!}\sum_{i_1,\dots,i_N}\int_0^tdt_1\dots \int_0^tdt_N \notag\\
&\quad\quad\quad\quad\times\sigma_{i}^{i_1\dots i_N}(t-t_1,\dots, t-t_N)\prod_{\ell=1}^NE_{i_\ell}(t_\ell).
\label{defsigma}
\end{align}
Here, $N$ denotes the order of the response, i.e., $N=1$ for the linear conductivity and $N\geq2$ for non-linear conductivities.  Summations of $i_\ell$'s ($\ell=1,\dots,N$) run over $x,y,\dots$.  The response function $\sigma_{i}^{i_1\dots i_N}(t_1,\dots, t_N)$ vanishes whenever $t_\ell<0$ for any $\ell=1,2,\dots,N$ due to the causality.  It is also symmetric with respect to the permutation of any pair of $(i_\ell,t_\ell)$ and $(i_{\ell'},t_{\ell'})$.

The Fourier transformation of $\sigma_{i}^{i_1\dots i_N}(t_1,\dots, t_N)$ is defined as
\begin{align}
&\sigma_i^{i_1\dots i_N}(\omega_1,\dots,\omega_N)\notag\\
&=\int_{0}^{\infty}dt_1\dots \int_{0}^{\infty}dt_N\sigma_i^{i_1\dots i_N}(t_1,\dots,t_N)\prod_{\ell=1}^Ne^{(i\omega_\ell-\eta)t_\ell}.
\label{deffourier}
\end{align}
The most singular part of $\sigma_i^{i_1\dots i_N}(\omega_1,\dots,\omega_N)$ around $\omega_1=\dots=\omega_N=0$ takes the form
\begin{align}
\sigma_{i\,\text{(Drude)}}^{i_1\dots i_N}(\omega_1,\dots,\omega_N)=\mathcal{D}_{i}^{i_1\dots i_N}\prod_{\ell=1}^N\frac{i}{\omega_\ell+i\eta}.\label{Drude}
\end{align}
We call $\mathcal{D}_{i}^{i_1\dots i_N}$ nonlinear Drude weight for $N\geq2$. The formula $(\omega+i\eta)^{-1}=\mathcal{P}\omega^{-1}-i\pi\delta(\omega)$ implies that this term contains $\prod_{\ell=1}^N\delta(\omega_\ell)$.
In real time, the Drude weight part of the conductivity reads
\begin{equation}
\sigma_{i\,\text{(Drude)}}^{i_1\dots i_N}(t_1,\dots,t_N)=\mathcal{D}_i^{i_1\dots i_N}\prod_{\ell=1}^N\theta(t_\ell).\label{Drudet}
\end{equation}
Here $\theta(t)$ is the step function. Note that the non-linear conductivity may contain other, more moderately singular terms. For example, $\sigma_{i}^{i_1i_2}(\omega_1,\omega_2)$ may contain $\delta(\omega_1)g(\omega_2)$ where $g(\omega_2)$ is regular around $\omega_2=0$.

\subsection{Main results}

The first main result of this work is the generalized $f$-sum rules of nonlinear conductivities:
\begin{align}
&\int_{-\infty}^{\infty} \frac{d\omega_1}{2\pi}\dots\int_{-\infty}^{\infty} \frac{d\omega_N}{2\pi}  \sigma_i^{i_1\dots i_N}(\omega_1,\dots,\omega_N)\notag\\
&=\frac{1}{2^NV}\Big\langle\frac{\partial^{N+1}\hat{H}(\vec{A})}{\partial A_{i}\partial A_{i_1}\dots \partial A_{i_N}}\Big|_{\vec{A}=\vec{0}}\Big\rangle_{0}.
\label{main1}
\end{align}
Here $\langle\hat{O}\rangle_0\equiv\text{tr}[\hat{O}\hat{\rho}(0)]$ is the expectation value defined by the unperturbed density matrix in Eq.~\eqref{initial}.
Any density-density interactions, or more generally any terms in Hamiltonian which do not couple to the gauge field, do not appear explicitly in the right-hand side of the $f$-sum rule.
The derivative of the Hamiltonian in this expression represents the explicit dependence of the current operator \eqref{defj} on $\vec{A}$, which is usually referred to as  the ``diamagnetic'' contribution.
Different types of $f$-sum rules of nonlinear conductivities have been discussed previously, for example, in Refs.~\cite{PhysRevB.44.8446,doi:10.1063/1.470283,PhysRevB.98.165113}.

The second main result is the generalized Kohn formula for nonlinear Drude weights:
\begin{align}
&\mathcal{D}_i^{i_1\dots i_N}=\frac{1}{V}\frac{\partial^{N+1}\mathcal{E}(\vec{A})}{\partial A_{i}\partial A_{i_1}\dots \partial A_{i_N}}\Big|_{\vec{A}=\vec{0}},\label{main2}\\
&\mathcal{E}(\vec{A})\equiv\sum_n \rho_n\mathcal{E}_n(\vec{A}).\label{main22}
\end{align}
Here, $\mathcal{E}_n(\vec{A})$ is the energy eigenvalue of the (instantaneous) eigenstate $|n(\vec{A})\rangle$ of $\hat{H}(\vec{A})$, which is assumed to be continuously connected to $|n(\vec{0})\rangle$. Level crossings may occur at a finite $\vec{A}$ and $\mathcal{E}_n(\vec{A})$ does not necessarily coincide with the $n$-th energy level of $\hat{H}(\vec{A})$. Note that, in general, $\mathcal{E}(\vec{A})$ cannot be interpreted as any sort of free energies as the weight $\rho_n$ is fixed independent of $\vec{A}$.
For noninteracting Bloch electrons in a periodic lattice, Ref.~\onlinecite{Parker2019} found an expression equivalent to Eq.~\eqref{main2} from a diagrammatic approach up to $N=3$ in the semi-classical limit.  Our result is much more general, being applicable to general
interacting systems and up to the infinite order.
The similarity between the generalized $f$-sum rule~\eqref{main1} and
the generalized Kohn formula~\eqref{main2} is now evident.
Yet, they are different, and the difference reflects the different
underlying processes, as we will discuss details in Sec.~\ref{sec:proof1}.
The generalized $f$-sum rule is given by the expectation value of the derivative
of the Hamiltonian, which corresponds to the quench process.
In contrast, the generalized Kohn formula is given by the derivative of
the energy eigenvalues, which corresponds to the adiabatic process.

Our results reproduce the well-known $f$-sum rule~\cite{Resta_2018} and the Kohn formula~\cite{Kohn1964,PhysRevLett.74.972,Resta_2018} for the linear conductivity. We also have an infinite series of generalized formulas for nonlinear conductivities.
Examples of second-order relations are
\begin{align}
&\int_{-\infty}^\infty  \frac{d\omega_1}{2\pi}\int_{-\infty}^\infty  \frac{d\omega_2}{2\pi} \sigma_{x}^{xx}(\omega_1,\omega_2)=\frac{1}{4V}\Big\langle\frac{\partial^3\hat{H}(\vec{A})}{\partial A_x^3}\Big|_{\vec{A}=\vec{0}}\Big\rangle_0,\\
&\int_{-\infty}^\infty  \frac{d\omega_1}{2\pi}\int_{-\infty}^\infty  \frac{d\omega_2}{2\pi} \sigma_{x}^{xy}(\omega_1,\omega_2)\notag\\
&=\int_{-\infty}^\infty  \frac{d\omega_1}{2\pi}\int_{-\infty}^\infty  \frac{d\omega_2}{2\pi} \sigma_{y}^{xx}(\omega_1,\omega_2)=\frac{1}{4V}\Big\langle\frac{\partial^3\hat{H}(\vec{A})}{\partial A_x^2\partial A_y}\Big|_{\vec{A}=\vec{0}}\Big\rangle_0\label{relsp1}
\end{align}
and
\begin{align}
&\mathcal{D}_{x}^{xx}=\frac{1}{V}\frac{\partial^{3}\mathcal{E}(\vec{A})}{\partial A_x^3}\Big|_{\vec{A}=\vec{0}},\\
&\mathcal{D}_{x}^{yz}=\mathcal{D}_{y}^{zx}=\mathcal{D}_{z}^{xy}=\frac{1}{V}\frac{\partial^3\mathcal{E}(\vec{A})}{\partial A_x\partial A_y\partial A_z}\Big|_{\vec{A}=\vec{0}}.\label{relsp2}
\end{align}
In particular, Eqs.~\eqref{relsp1} and \eqref{relsp2} imply unexpected relations among distinct components of nonlinear conductivities in different spatial directions. We stress that they are derived without assuming any spatial symmetry.

The order-by-order expression of the Drude weights \eqref{main2} can be combined together into a compact form that fully contains the effect of $\vec{A}(t)$ to all orders.
\begin{align}
j_{i\,\text{(Drude)}}(t)=\frac{1}{V}\frac{\partial\mathcal{E}(\vec{A})}{\partial A_{i}}\Big|_{\vec{A}=\vec{A}(t)}.
\label{Bloch1}
\end{align}
Here, $j_{i\,\text{(Drude)}}(t)$ is the part of $j_i(t)$ including all contributions from the linear and nonlinear Drude weights.

Under the open boundary condition, the effect of nonzero $\vec{A}$ can be ``gauged away" to outside of the system. Hence, the energy eigenvalue $\mathcal{E}_n(\vec{A})$ cannot actually depends on $\vec{A}$ and the Drude weight vanishes at all orders. This is consistent with the previous study~\cite{RigolShastry-Drude} which found the vanishing linear Drude weight under the open boundary condition.

When the periodic boundary condition with the period $L_i$ in the $i$-th direction is instead imposed, the gauge field $A_i$ can be interpreted as the twist $\phi_i=A_i L_i$ of the boundary condition.  Although the Hamiltonian $\hat{H}(\vec{A})$ with $\phi_i=2\pi n_i$ ($n_i\in\mathbb{Z}$) is unitary equivalent to $\hat{H}(\vec{0})$, this does not necessarily imply $\mathcal{E}_n(\vec{A})=\mathcal{E}_n(\vec{0})$ because of the possible level crossings
remarked above~\cite{PhysRevLett.65.1833,Oshikawa-Drude,*Oshikawa-Drude-Erratum}.

\section{Derivation of the Main Results}
\label{sec:proof1}

We derive our formulas by considering a time-evolution process where $A_i(t)$ is increased from $0$ at $t=0$ to a constant $\mathcal{A}_i$ at $t=T$. To precisely formulate this process, let us write
\begin{equation}
A_i(t)=f_i(t/T)\mathcal{A}_i,
\end{equation}
where $f_i(\tau)$ is an analytic function of $\tau\in\mathbb{R}$, satisfying $f_i(\tau)=0$ for $\tau\leq0$ and $f_i(\tau)=1$ for $\tau\geq1$.  It is crucial that the value of $A_i(T)=\mathcal{A}_i$ is fixed independent of $T$.

The common strategy in our discussion of the generalized $f$-sum rule and Kohn formula is to evaluate  $j_i(T)=\langle\hat{j}_i(\mathcal{\vec{A}})\rangle_T$ in two different ways, one directly from Eqs.~\eqref{rhot} and \eqref{defj2} and the other using Eq.~\eqref{defsigma}.  We then compare the coefficient of $\prod_{\ell=1}^N\mathcal{A}_{i_\ell}$ in the two expressions and derive constraints.

\subsection{$f$-sum rule}
We start with the $f$-sum rule. To this end, we consider the limit of very quick change of the vector potential: $T \to 0$. This can be regarded as an example of quantum quench
(sudden switching of the vector potential). In this limit, the state cannot follow the change of the Hamiltonian, and
``the sudden approximation $\hat{S}(T)=1$'' becomes exact. This can be most easily seen by the formula ($\calT$ denotes the time-ordering)
\begin{align}
\hat{S}(T)=\mathcal{T} e^{ -i T\int_0^1 d\tau\hat{H}(f_i(\tau)\mathcal{A}_i)}.
\end{align}
Because of the prefactor $T$ in the exponent, $\hat{S}(T)\to 1$ in the limit of $T\to0$.  In this limit, all responses of the electric current originate from the diamagnetic contributions.

Let us evaluate $j_i(T)=\langle\hat{j}_i(\mathcal{\vec{A}})\rangle_T$ in two different ways. On the one hand, $\langle\hat{O}\rangle_T$ can be approximated by $\langle\hat{O}\rangle_0$ in the quench limit. Thus
\begin{align}
&j_i(T)=\frac{1}{V}\Big\langle\frac{\partial \hat{H}(\vec{A})}{\partial A_i}\Big|_{\vec{A}=\vec{\mathcal{A}}}\Big\rangle_0\notag\\
&=\sum_{N=0}^\infty\frac{1}{N!V}\sum_{i,i_1\dots i_N}\Big\langle\frac{\partial^{N+1}\hat{H}(\vec{A})}{\partial A_{i}\partial A_{i_1}\dots \partial A_{i_N}}\Big|_{\vec{A}=\vec{0}}\Big\rangle_0\prod_{\ell=1}^N\mathcal{A}_{i_\ell}.\label{derivation1}
\end{align}
On the other hand, when  $T$ is small enough, $\sigma_{i}^{i_1\dots i_N}(t-t_1,\dots, t-t_N)$ in Eq.~\eqref{defsigma} can be approximated by
\begin{align}
\sigma_{i}^{i_1\dots i_N}(0)\equiv\lim_{t_1,\dots,t_N\to +0}\sigma_{i}^{i_1\dots i_N}(t_1,\dots, t_N).\label{inst}
\end{align}
We can then easily perform all the $\int_0^tdt_\ell$ integrals in Eq.~\eqref{defsigma}  and get
\begin{align}
j_i(T)-j_i(0)=\sum_{N=1}^\infty\frac{1}{N!}\sum_{i_1\dots i_N} \sigma_{i}^{i_1\dots i_N}(0)\prod_{\ell=1}^N\mathcal{A}_{i_\ell}.\label{derivation3}
\end{align}
Comparing Eqs.~\eqref{derivation1} and \eqref{derivation3}, we find
\begin{align}
\sigma_{i}^{i_1\dots i_N}(0)=\frac{1}{V}\Big\langle\frac{\partial^{N+1}\hat{H}(\vec{A})}{\partial A_{i}\partial A_{i_1}\dots \partial A_{i_N}}\Big|_{\vec{A}=\vec{0}}\Big\rangle_0.
\label{instantaneous}
\end{align}
Finally, this relation can be cast into the form of $f$-sum rules \eqref{main2} by expressing $\sigma_{i}^{i_1\dots i_N}(0)$ in terms of the Fourier component.
\begin{align}
\int_{-\infty}^{\infty} \frac{d\omega_1}{2\pi}\dots\int_{-\infty}^{\infty} \frac{d\omega_N}{2\pi}  \sigma_i^{i_1\dots i_N}(\omega_1,\dots,\omega_N)=\frac{\sigma_{i}^{i_1\dots i_N}(0)}{2^N}.\label{derivation2}
\end{align}
The factor $2^{-N}$ originates from the discontinuity of $\sigma_{i}^{i_1\dots i_N}(t_1,\dots,t_N)$ around $t_{\ell}=0$.

\subsection{Kohn formula}
Let us move onto the Kohn formula.  This time we consider the opposite limit; that is, the limit of the adiabatic flux insertion, $T \to \infty$.~\cite{Oshikawa-Drude,*Oshikawa-Drude-Erratum}
In this limit, the adiabatic theorem~\cite{doi:10.1143/JPSJ.5.435,2002.02947} tells us that $\hat{S}(T)|n(\vec{0})\rangle\propto |n(\vec{\mathcal{A}})\rangle$ so that
\begin{align}
\hat{\rho}(T)=\sum_n\rho_n|n(\vec{\mathcal{A}})\rangle\langle n(\vec{\mathcal{A}})|.
\end{align}
Crucially, the weight $\rho_n$ remains unchanged even when energy levels $\mathcal{E}_n(\vec{\mathcal{A}})$ explicitly depend on $\vec{\mathcal{A}}$.  Thus using the Hellmann--Feynman theorem, we find
\begin{align}
j_i(T)&=\frac{1}{V}\sum_n\rho_n\Big\langle n(\vec{A})\Big|\frac{\partial \hat{H}(\vec{A})}{\partial A_i}\Big|n(\vec{A})\Big\rangle\Big|_{\vec{A}=\vec{\mathcal{A}}}\notag\\
&=\frac{1}{V}\sum_n\rho_n\frac{\partial \mathcal{E}_n(\vec{A})}{\partial A_i}\Big|_{\vec{A}=\vec{\mathcal{A}}}=\frac{1}{V}\frac{\partial \mathcal{E}(\vec{A})}{\partial A_i}\Big|_{\vec{A}=\vec{\mathcal{A}}}\notag\\
&=\sum_{N=1}^\infty\frac{1}{N!V}\sum_{i_1\dots i_N}\frac{\partial^{N+1}\mathcal{E}(\vec{A})}{\partial A_{i}\partial A_{i_1}\dots \partial A_{i_N}}\Big|_{\vec{A}=\vec{0}}\prod_{\ell=1}^N\mathcal{A}_{i_\ell}.
\label{derivation4}
\end{align}

Next we show that only the Drude weight contribution is important for the current response in the adiabatic limit. To this end, let us use the Fourier transformation and rewrite the right-hand side of Eq.~\eqref{defsigma} as
\begin{align}
&\sum_{N=1}^\infty\frac{1}{N!}\sum_{i_1,\dots,i_N}\prod_{\ell=1}^{N}\mathcal{A}_{i_\ell}\int_{-\infty}^{\infty} \frac{d\omega_1}{2\pi}\dots\int_{-\infty}^{\infty} \frac{d\omega_N}{2\pi}\notag\\
&\quad\quad\quad\quad\quad\quad\quad\times \sigma_{i}^{i_1\dots i_N}(\omega_1,\dots, \omega_N) \prod_{\ell=1}^NI_{i_\ell}(\omega_{\ell}),\label{integral}
\end{align}
where
\begin{align}
I_{i}(\omega)\equiv\int_0^1 d\tau e^{i\omega T(\tau-1)} \frac{df_{i}(\tau)}{d\tau}.
\end{align}
When $\omega=0$,  $I_{i}(0)=f_i(1)=1$. However, when $\omega\neq0$, we can derive the following upper-bound using an integration by part and the Schwartz inequality:
\begin{align}
\left|I_{i}(\omega)\right|=\left|\int_0^1 d\tau \frac{1}{i\omega T}\frac{de^{i\omega T(\tau-1)}}{d\tau} \frac{df_{i}(\tau)}{d\tau}\right|\leq\frac{C_{i}}{|\omega| T},
\end{align}
where $C_{i}\equiv\max_{0\leq\tau\leq1}\big(2|df_i(\tau)/d\tau|+|d^2f_i(\tau)/d\tau^2|\big)$ is a finite constant because of the assumed analyticity of $f_i(\tau)$. Thus $\lim_{T\to\infty}I_{i}(\omega)=0$ when $\omega\neq0$.  This means that only the term proportional to $\prod_{\ell=1}^N\delta(\omega_\ell)$ in $\sigma_{i}^{i_1\dots i_N}(\omega_1,\dots, \omega_N)$, i.e., the Drude weight term \eqref{Drude}, can contribute to the integral in Eq.~\eqref{integral} in the adiabatic limit.

Finally, the contribution from the Drude weight in the current response can be readily evaluated by plugging Eq.~\eqref{Drudet} into Eq.~\eqref{defsigma}:
\begin{align}
j_i(T)-j_i(0)=\sum_{N=1}^\infty\frac{1}{N!}\sum_{i_1\dots i_N} \mathcal{D}_{i}^{i_1\dots i_N}\prod_{\ell=1}^N\mathcal{A}_{i_\ell}.
\label{derivation5}
\end{align}
Comparing the coefficient of $\prod_{\ell=1}^N\mathcal{A}_{i_\ell}$ between Eqs.~\eqref{derivation4} and \eqref{derivation5}, we obtain the generalized Kohn formula \eqref{main2}.

\section{Examples}
\label{example}
\subsection{Tight-binding models}
Let us clarify the physical implication of the nonlinear Drude weights by considering noninteracting electrons subjected to a periodic potential.  Suppose that a constant electric field $\vec{E}$ is applied to this system at a finite temperature. If we assume the periodic boundary condition, Eq.~\eqref{Bloch1} for this setting becomes
\begin{align}
j_{i\,\text{(Drude)}}(t)=\frac{1}{V}\sum_{\alpha,\vec{k}}n(\varepsilon_{\alpha,\vec{k}})\partial_{k_i}\varepsilon_{\alpha,\vec{k}+\vec{E}t},
\label{Bloch2}
\end{align}
where $\vec{k}$ is the crystal momentum, $\varepsilon_{\alpha,\vec{k}}$ is the band dispersion of $\alpha$-th band, and $n(\varepsilon)\equiv1/(e^{\beta\varepsilon}+1)$ is the Fermi--Dirac distribution function. Thus electrons under a periodic potential, in general, exhibit nonlinear responses toward the applied electric field unless they form a band insulator. This is in sharp contrast to electrons in free space which are simply accelerated at the constant rate $\vec{E}/m_{\text{el}}$ ($m_{\text{el}}$ is the electron mass).  Because the band dispersion $\varepsilon_{\alpha,\vec{k}}$ is periodic in $\vec{k}$,  Eq.~\eqref{Bloch2} implies that electrons will go back and forth. This is nothing but the well-known Bloch oscillation~\cite{LEO1992943,PhysRevB.51.17275,PhysRevLett.76.4508,Hartmann_2004}.

To give a simple example in which $\mathcal{E}(\vec{A})$ in Eq.~\eqref{main22} has a nontrivial $\vec{A}$-dependence even at a finite temperature, let us discuss the $d=1$ tight-binding model with a nearest neighbor hopping $t>0$ at half filling:
\begin{align}
\hat{H}(A_x)&=-t\sum_{x=1}^{L_x}(\hat{c}_{x+1}^\dagger e^{-iA_x}\hat{c}_{x}+\text{h.c.})\notag\\
&=\sum_{k_x}\varepsilon_{k_x+A_x}\hat{c}_{k_x}^\dagger \hat{c}_{k_x},\label{model}
\end{align}
Here, the lattice constant is set to be $1$, the band dispersion is given by $\varepsilon_{k_x}=-2t\cos k_x$, and the Fourier transformation is defined as $\hat{c}_{x}^\dagger=L_x^{-1/2}\sum_{k_x}e^{-ik_x x}\hat{c}_{k_x}^\dagger$.  Since $\varepsilon_{k_x}$ has a particularly simple form, the $A_x$-dependence of $\mathcal{E}(A_x)= \sum_{k_x}n(\varepsilon_{k_x})\varepsilon_{k_x+A_x}$ can be easily factored out:
\begin{equation}
\mathcal{E}(A_x)= \langle\hat{H}(0)\rangle_0\cos A_x.
\label{eq.E_Ax_1dTB}
\end{equation}
In fact since the Bloch function lacks the $A_x$-dependence in this one-band model, we have
\begin{equation}
\langle\hat{H}(A_x)\rangle_0=\mathcal{E}(A_x).
\label{eq.H_eq_E_1dTB}
\end{equation}
Therefore, the non-linear Drude weight agrees exactly with the $f$-sum
at the same order.
In other words, in this one-band tight-binding model,
the induced current does not depend on the timescale of the
application of the electric field, and is the same for the instantaneous
or adiabatic process.

Moreover, the simple functional form of Eq.~\eqref{eq.E_Ax_1dTB}
implies that,
the non-linear $f$-sum or the nonlinear Drude weight of all odd orders have the same amplitude in this model. The Drude weight at every even order vanishes due to the time-reversal symmetry.  The energy density $\langle\hat{H}(0)\rangle_0/L_x$ in the large $L_x$ limit changes continuously from $-(2t/\pi)[1-(\pi^2/24)(\beta t)^{-2}+O((\beta t)^{-4})]$ at low temperatures ($\beta t\gg1$) and $-(t/2)[\beta t+O((\beta t)^3)]$ at high-temperatures ($\beta t\ll1$).

\begin{figure*}[t]
\begin{center}
\includegraphics[width=0.70\textwidth]{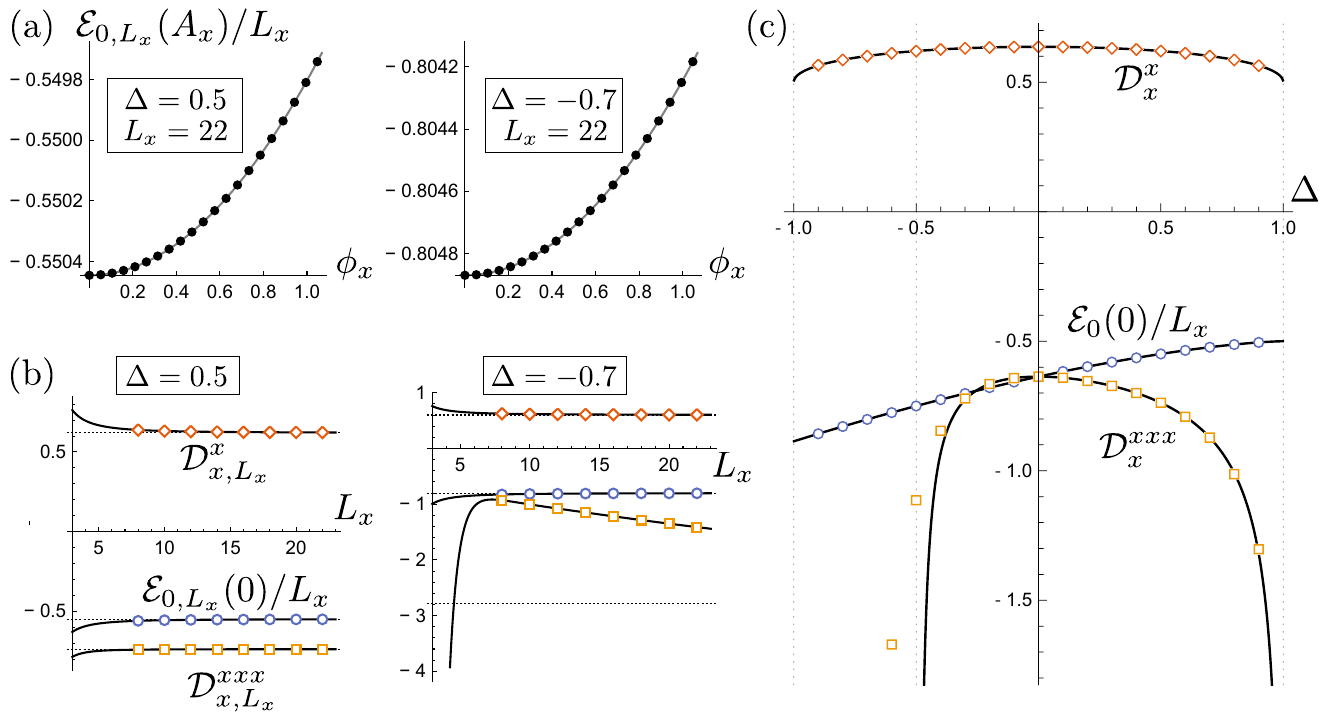}
\caption{\label{fig}  Numerical results for the $S=1/2$ XXZ chain at zero temperature, obtained by the exact diagonalization up to $L_x=22$ spins. All vertical axes are scaled with $J/2$.
(a) The ground state energy density as a function of $\phi_x=A_x L_x$.  The gray fitting curve is obtained by assuming Eq.~\eqref{fit1}.
(b) Extrapolation of the large $L_x$ values using the data for $L_x=8, 10,\dots, 22$.
(c) The ground state energy density $\mathcal{E}_{0}(0)/L_x$, the linear Drude weight $\mathcal{D}_x^x$, and the third-order Drude weight $\mathcal{D}_x^{xxx}$ in the large $L_x$ limit as a function of $\Delta$. The black curves represent analytic results of the ground state energy density~\cite{PhysRev.150.327}, the linear Drude weight~\cite{PhysRevLett.65.1833}, and the third order Drude weight (thermodynamic limit of the result in Ref.~\cite{LUKYANOV1998533}).}
\end{center} 
\end{figure*}

\subsection{$S=1/2$ XXZ chain}

Finally, as an example of interacting models, let us discuss the $S=1/2$ anisotropic Heisenberg spin chain ($J>0$) at  zero temperature:
\begin{equation}
\hat{H}(A_x)=-J\sum_{x=1}^{L_x}\Big(\frac{1}{2}\hat{s}_{x+1}^+e^{-iA_x}\hat{s}_x^-+\text{h.c.}+\Delta \hat{s}_{x+1}^z\hat{s}_{x}^z\Big).
\label{xxz}
\end{equation}
Again we assume the periodic boundary condition.

The $\Delta=0$ case reduces to the tight-binding model \eqref{model} with $t=J/2$.
As we have discussed in the previous subsection, in this case,
Eqs.~\eqref{eq.E_Ax_1dTB} and~\eqref{eq.H_eq_E_1dTB} implies
that the linear Drude weight $\mathcal{D}_x^x$ coincides with the linear $f$-sum,
the second-order Drude weight and $f$-sum vanish,
and the third-order Drude weight is given by
\begin{equation}
\mathcal{D}_x^{xxx} = - \mathcal{D}^x_x
\label{eq.Dx3_vs_Dx}
\end{equation}
which coincides with the third-order $f$-sum.

We can now see the effect of interactions by turning to $\Delta \neq 0$.
An analytic expression of the linear Drude weight $\mathcal{D}_x^x$ in the large $L_x$ limit was obtained~\cite{PhysRevLett.65.1833} by applying the Kohn formula to the results~\cite{Hamer_1987} of Bethe ansatz. In our notation, it reads
\begin{equation}
\mathcal{D}_x^x=\frac{\pi J}{4}\frac{\sin\gamma}{\gamma(\pi-\gamma)}
\label{xxz2}
\end{equation}
for $\Delta=-\cos\gamma$ ($0\leq\gamma<\pi$).

To calculate the third-order Drude weight $\mathcal{D}_x^{xxx}$ for $|\Delta|<1$, we perform the exact diagonalization up to $L_x=22$ spins.
For each $\Delta$, we compute the ground state energy $\mathcal{E}_{0,L_x}(A_x)$ as a function of $A_x$ [Fig.~\ref{fig} (a)] and determine $\mathcal{D}_{x,L_x}^{xxx}$ by assuming the Taylor series of the form
\begin{equation}
\frac{\mathcal{E}_{0,L_x}(A_x)}{L_x}=\frac{\mathcal{E}_{0,L_x}(0)}{L_x}+\frac{\mathcal{D}_{x,L_x}^xA_x^2}{2}+\frac{\mathcal{D}_{x,L_x}^{xxx}A_x^4}{24}+O(A_x^6).
\label{fit1}
\end{equation}
We note that, for the given system size $L_x$, the Drude weight at each order is well-defined and obeys the generalized Kohn formula~\eqref{main2}.
In the actual calculation, we use $\phi_x\equiv A_xL_x$ in the range $0\leq \phi_x \leq \pi/3$, limiting $A_x$ to be small enough to avoid any level crossings.  To check the accuracy of this part of our calculation, we compare the values of $\mathcal{D}_{x,L_x}^{xxx}$ obtained this way with an independent calculation via Kubo's response theory~\cite{derivation} that does not involve a gauge field for $L_x=4,6,\dots,14$.  We found that the error was less than $10^{-7}$ for all $\Delta$.

We repeat this calculation for $L_x=8, 10,\dots, 22$ and
 estimate the values in the thermodynamic ($L_x \to \infty$) limit assuming the power-law decay $\mathcal{D}_{x,L_x}^{xxx}=\mathcal{D}_x^{xxx}+\sum_{m=1}^4c_mL_x^{-m}$.
The extrapolation works well for $\Delta\gtrsim -0.3$ [see the left panel of Fig.~\ref{fig} (b)], while it fails for $\Delta\lesssim -0.3$ [the right panel of Fig.~\ref{fig} (b)].
In fact, for $-1/2<\Delta<1$, we find an exact analytic expression
\begin{equation}
\mathcal{D}_x^{xxx}=-\frac{J\sin\gamma}{16\gamma(\pi-\gamma)}\biggl(
\frac{
\Gamma\big(\frac{3\pi}{2\gamma}\big)\Gamma\big(\frac{\pi-\gamma}{2\gamma}\big)^3
}{
\Gamma\big(\frac{3(\pi-\gamma)}{2\gamma}\big)\Gamma\big(\frac{\pi}{2\gamma}\big)^3
}+\frac{3\pi\tan\big(\frac{\pi^2}{2\gamma}\big)}{\pi-\gamma}\biggr)
\label{xxxx}
\end{equation}
by taking the thermodynamic limit of the result based on an effective field theory in Ref.~\cite{LUKYANOV1998533}, where $\Gamma(z)$ is the gamma function.

We find that the non-linear Drude weight $\mathcal{D}_x^{xxx}$ has a nontrivial dependence on the interaction $\Delta$, as shown in Fig.~\ref{fig} (c).
To verify our calculation, we also perform the same analysis for the ground state energy density $\mathcal{E}_{0}(0)/L_x$ and estimate the linear Drude weight $\mathcal{D}_{x}^x$ in the thermodynamic limit. As seen in the upper panel of the Fig.~\ref{fig} (c), the obtained result shows an excellent agreement with the known analytic results~\cite{PhysRev.150.327,PhysRev.150.327} in the entire parameter range $-1 \leq \Delta < 1$. This supports the reliability of our numerical calculation.
The nonlinear Drude weight $\mathcal{D}_x^{xxx}$ obtained numerically as described above also shows a good agreement with the exact analytic formula, especially in the region $\Delta \gtrsim -0.3$ where the extraporation to the thermodynamic limit works well. On the other hand, the numerical result shows some deviation from the exact formula as $\Delta$ approaches $-1/2$ from the above. This presumably reflects the divergence of $\mathcal{D}_x^{xxx}$ in the limit $\Delta \to -1/2+0$ and the small system size used in the numerical diagonalization. Considering this, the numerical result is qualitatively consistent with the analytic formula in the range $-1/2 < \Delta \lesssim -0.3$.
In fact, within the effective field theory approach,  $\mathcal{D}_x^{xxx}$ diverges in the the thermodynamic limit for the entire range of $-1\leq\Delta\leq-1/2$, and this behavior is also supported by our numerical result.
We leave for the future work further investigation of the mechanism and the physical implication of the divergent behavior of $\mathcal{D}_{x}^{xxx}$ for $-1\leq\Delta\leq-1/2$.

We note that, for the present model,
\begin{align}
\frac{\partial^{2m}\hat{H}(A_x)}{\partial A_x^{2m}}\Big|_{A_x=0}=(-1)^{m-1}\frac{\partial^{2}\hat{H}(A_x)}{\partial A_x^{2}}\Big|_{A_x=0},\\
\frac{\partial^{2m-1}\hat{H}(A_x)}{\partial A_x^{2m-1}}\Big|_{A_x=0}=(-1)^{m-1}\frac{\partial\hat{H}(A_x)}{\partial A_x}\Big|_{A_x=0}
\end{align}
for $m\geq1$. Therefore, the right-hand side of the $f$-sum rule at all odd orders have the same magnitude with the alternating sign, and that of all even orders vanish.
In contrast, Fig.~\ref{fig} (c) clearly shows that the linear and third-order Drude weights are generally different.
The simple relation~\eqref{eq.Dx3_vs_Dx}, which was derived for the
non-interacting tight-binding model, breaks down once the interaction is included ($\Delta \neq 0$.)

\section{discussions}
\label{sec:discussion}
In this work, we obtained an infinite series of new $f$-sum rules \eqref{main1} and Kohn formulas \eqref{main2} on the
nonlinear conductivities.  We found nontrivial relations among conductivities in different spatial directions, such as  Eqs.~\eqref{relsp1} and \eqref{relsp2},
even in the absence of any spatial symmetry.

In the discussion of the nonlinear $f$-sum rules, we did not use the explicit form of the initial state $\hat{\rho}(0)$ given in Eq.~\eqref{initial}. In fact, $\hat{\rho}(0)$ can be chosen to be a non-equilibrium state~\cite{ShimizuYuge1,ShimizuYuge2,RevModPhys.86.779}, especially a non-equilibrium steady state for which the response function would still be time-translation invariant. For a more general non-equilibrium state, where the response function lacks the time-translation invariance, the $f$-sum rule should be understood as the constraint on the instantaneous conductivity~\cite{derivation}.

The nonlinear $f$-sum rules can also be extended to position-dependent responses toward non-uniform electric fields on an arbitrary lattice.
To see this, let $L$ be the set of directed links (arrows), each of which connects a pair of lattice sites.
The local vector potential $A_{l}(t)$ on each link $l\in L$, and hence the local electric field $E_{l}(t)\equiv dA_{l}(t)/dt$, are allowed to depend on $l$.
We are interested in the response of the local current density, defined by $\hat{j}_{l}(t)\equiv\partial \hat{H}(t)/\partial A_{l}(t)$ for each link, towards the position-dependent electric field $E_{l'}(t)$. One can simply re-use all of our discussions in this work without any formal change by replacing $i$'s (indices for spatial directions) with $l$'s (indices for links).  In general, the position-dependent vector potentials $A_{l}(t)$ may also produce a local magnetic field and Eq.~\eqref{defsigma} needs to be modified. However, the effect of such magnetic fields is suppressed by a factor of $T$ (duration of the time evolution) and can be neglected in the quench limit $T\rightarrow0$ relevant for the instantaneous response.

While we used lattice models in our derivation, essentially the same argument applies to continuum models as well.  For the particular case of the non-relativistic quantum mechanical
Hamiltonian $
\hat{K} = \int d^dr\,\hat{c}^\dagger_{\vec{r}}
\big[- (\vec{\nabla}^2/2m) - \mu \big]
\hat{c}_{\vec{r}},
$ with density-density interaction,
the right-hand side of the $f$-sum rule vanishes
for all nonlinear conductivities.
Although this is rather remarkable, this does not imply the absence
of any nonlinear response to the electric field.
The vanishment of the $f$-sum rule just implies that
any positive part of $\sigma_{i}^{(i_1,\dots,i_N)}(\omega_1,\dots,\omega_N)$
must be compensated by a negative part.

Since the lattice models for electron systems are low-energy effective
model for non-relativistic electrons in crystal, the nonlinear
$f$-sum of a real electron system
would vanish by integrating over the infinite frequency range.
A non-vanishing $f$-sum for the low-energy lattice model should
correspond to an frequency integral up to the cutoff energy,
typically the order of the bandwidth of the lattice model.

A non-vanishing $f$-sum rule for a low-energy
effective model at a given order $N$
does indicate the presence of the $N$-th order conductivity.
While the maximum of the desired $N$-th order effect, such as the
shift current at $N=2$, would be generally different from the maximum
of the $f$-sum at the same order, the latter is easier to evaluate
and could give a quick guidance for construction of a model with
a desired property (such as a large shift current).

The present result is one of rather
few general constraints on conductivities, especially non-linear ones.  The sum rules can be used to check various approximations or numerical
calculations, and would give a guiding principle on designing
systems with  desired transport properties.
We hope that the present result will help developing theory
of linear and nonlinear dynamical responses of
quantum many-body systems in the future.

\begin{acknowledgments}
This work is initiated while M.~O. was participating in the Harvard CMSA Program on \textit{Topological Aspects of Condensed Matter}.
He thanks Yuan-Ming Lu, Ying Ran, and Xu Yang, for the discussions during the program which eventually led to the present work.
A part of the work by M.~O. was also performed at the Aspen Center for Physics, which is supported by National Science Foundation Grant PHY-1607611.
We thank Yoshiki Fukusumi, Naoto Nagaosa, Marcos Rigol, and Sriram Shastry for useful comments on the early version of the draft.
We are particularly grateful to Kazuaki Takasan for suggesting potential applications to non-equilibrium steady states and Takahiro Morimoto for pointing out the connection to the Bloch oscillation.
We also acknowledge useful discussions, including collaborations on related earlier projects, with
Yoshiki Fukusumi, Shunsuke C. Furuya, Ryohei Kobayashi,
Gr\'{e}goire Misguich, Yuya Nakagawa, and Masaaki Nakamura.
The work of M.O. was supported in part by MEXT/JSPS KAKENHI Grant Nos. JP19H01808 and JP17H06462, and JST CREST Grant Number JPMJCR19T2, Japan.
The work of H.W. is supported by JST PRESTO Grant No. JPMJPR18LA.
\end{acknowledgments}

\bibliography{bibs}

\end{document}